\documentclass[journal=jpclcd,manuscript=article]{achemso}
\usepackage[version=3]{mhchem} 
\usepackage{achemso}
\usepackage{graphicx} 
\usepackage{dcolumn} 
\usepackage{bm} 
\usepackage[utf8]{inputenc}
\usepackage[T1]{fontenc}  
\usepackage{hyperref} 
\usepackage{enumerate} 
\usepackage{float}
\usepackage{physics,color}
\usepackage{amsmath} 
\newcommand{\mycomment}[1]{\textcolor{black}{#1}}

\title{Modeling Spin-Dependent Nonadiabatic Dynamics with Electronic Degeneracy: A Phase-Space Surface-Hopping Method} 

\author{Xuezhi Bian}
\affiliation{Department of Chemistry, University of Pennsylvania, Philadelphia, Pennsylvania 19104, USA}
\author{Yanze Wu}
\affiliation{Department of Chemistry, University of Pennsylvania, Philadelphia, Pennsylvania 19104, USA}
\author{Jonathan Rawlinson}
\affiliation{Department of Mathematics, University of Manchester, Manchester M13 9PL, UK}
\author{Robert G. Littlejohn}
\affiliation{Department of Physics, University of California, Berkeley, California 94720, USA}
\author{Joseph E. Subotnik}
\affiliation{Department of Chemistry, University of Pennsylvania, Philadelphia, Pennsylvania 19104, USA}
\email{subotnik@sas.upenn.edu}

\begin{document}
\begin{tocentry} 
\includegraphics[width=8cm]{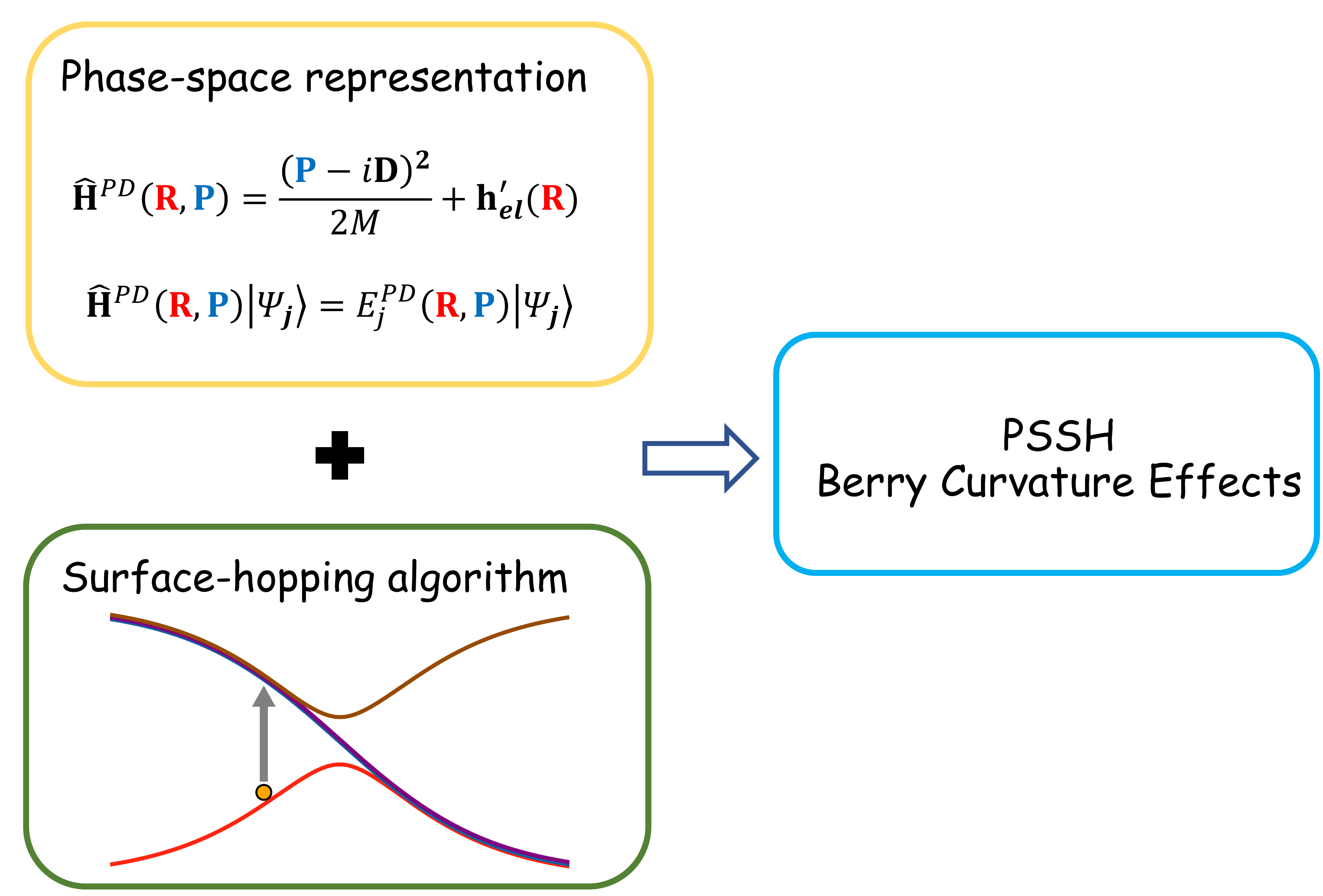}
\end{tocentry}

\begin{abstract}
Nuclear Berry curvature effects emerge from electronic spin degeneracy and can lead to non-trivial spin-dependent (nonadiabatic) nuclear dynamics. 
However, such  effects are completely neglected in all current mixed quantum-classical methods such as fewest switches surface-hopping.
In this work, we present a phase-space surface-hopping (PSSH) approach to simulate singlet-triplet intersystem crossing  dynamics. 
We show that with a simple pseudo-diabatic ansatz, a PSSH algorithm can capture the relevant Berry curvature effects and make predictions in agreement with exact quantum dynamics for a simple singlet-triplet model Hamiltonian. Thus, this approach represents an important step towards simulating photochemical and spin processes concomitantly,  as relevant to intersystem crossing and spin-lattice relaxation dynamics.

\end{abstract}

While not always fully appreciated, the spin degrees of freedom pertaining to a molecular or material system can be of particular importance when electronic transitions occur between states with different spin multiplicities (e.g. intersystem crossing [ISC])  or when states with large multiplicity interconvert  (e.g., triplet internal conversion [IC])\cite{Kasha1950,Penfold2018}. 
After all, nonadiabatic  nuclear-electronic dynamics underlie many key phenomena in physical chemistry and chemical physics, including scattering\cite{Roy2009}, charge transfer\cite{Akimov2013} and photochemical processes\cite{Tretiak2020}, and non-trivial spin degrees of freedom are almost always present.
For instance, recent experiments of water (\ce{H2O}) splitting have shown that, if one can produce \ce{OH} radicals of the same spin, one can effectively block \ce{H2O2} production and increase the yield of \ce{H2} and \ce{O2}\cite{Waldeck2020}.  
Currently, there are a wide range of mixed quantum-classical (MQC) methods\cite{Agostini2019, Barbatti2018,Coker1995, Tully1998} such as mean-field Ehrenfest dynamics(MFE) \cite{Schwartz2005,Cotton2013}, fewest switches surface hopping (FSSH)\cite{Tully1990, Wang2016, Subotnik2016} and {\em ab initio} multiple spawning (AIMS)\cite{Ben2000} for efficiently modeling {\em ab initio} nonadiabatic dynamics with reasonable accuracy, and many of these methods have been extended in principle to models of spin-crossover dynamics \cite{Varganov2021, Mai2015, Granucci2012, Varganov2016}. However, there are clear limitations to what physics these algorithms can capture. In particular, none of the algorithms above can model the effects of Berry curvature (and  Berry force) in full generality \cite{Bian2021}.

For the sake of completeness, a small introduction to Berry curvature would now appear appropriate. In principle, Berry curvature effects come in two flavors. The first flavor arises when a molecule with an even number of electrons is in a magnetic field and nuclear dynamics follow a single (adiabatic) electronic potential energy surface that is reasonably  separated energetically from all other surfaces\cite{Mead1979,Berry1984,Berry1993}.  
In such a case, as the result of non-Born Oppenheimer derivative couplings to the other electronic adiabatic surfaces, the nuclei will experience a Lorentz like force of the form:
\begin{eqnarray} \label{eq:bf}
{\mathbf{F}^B_{j}}  = \mathbf{\Omega}_j \cdot  \mathbf{\dot  R}    =    (i\hbar \nabla \times \mathbf{d}_{jj}) \cdot \mathbf{\dot R}  = \sum_k 2\hbar \Im \left(\mathbf{d}_{jk} \cdot \mathbf{\dot  R}\right)   \mathbf{d}_{kj}
\end{eqnarray}
Here \mycomment{${\mathbf{F}^B_{j}}$ and} $\mathbf{\Omega}_j $ denote \mycomment{the Berry force and} Berry curvature on surface $j$\mycomment{, $\mathbf{\dot  R}$ is the nuclear velocity} and $\mathbf{d}_{jk}$ is the derivative coupling vector between adiabats $\ket{\psi_j}$ and $\ket{\psi_k}$. 
Note that, when spin-related couplings (spin-orbit couplings [SOC] \cite{Fedorov2003}, spin-spin couplings \cite{Higuchi1963}, magnetic fields \cite{Culpitt2021}, etc.) are taken into account, the system Hamiltonian and the derivative coupling become complex-valued
and the Berry force in Eq. \ref{eq:bf} is nonzero (and can be large) \cite{Wu2021ci}.

The second flavor of Berry curvature is the full nonadiabatic tensor that arises when multiple states come together and there is a crossing. For instance, consider a system with a number of degenerate or nearly degenerate states cross each other and one cannot separate the electronic manifold into a well defined set of electronic states that interact in a fairly simple pairwise fashion; a common example would be an ISC event where a singlet crosses three triplets and all three triplets are deeply entangled and interact with the singlet \cite{Bian2021}. In such a case,  Berry curvature effects arise even if the Hamiltonian is real-valued (with the on-diagonal terms of the Berry curvature tensor exactly zero). \mycomment{To our knowledge, effectively all of the quantum-classical schemes  that are popularly used today (as propagated either in the adiabatic or diabatic representations) do not include such Berry curvature effects in their non-Born Oppenheimer dynamics.}

To that end, over the last few years, we have been working intensively to find a path to incorporate Berry curvature effects into the FSSH algorithm (for both the non-degenerate and degenerate cases)\cite{Miao2019,Wu2021, Bian2022}. Our  algorithms (published in Refs. \citenum{Miao2019}-\citenum{Bian2022}) were based on the premise of including Berry forces when propagating along a given adiabatic state in the usual FSSH framework. The final routine was able to perform fairly well on a variety of model Hamiltonians, but with two severe drawbacks: $(i)$ the algorithms were very complicated and essentially {\em ad hoc} (beyond the usual assumptions/approximations/guesses inherent in FSSH\cite{Subotnik2013, kapral2016}), and  $(ii)$ the algorithms were hard to implement in {\em ab initio} calculations. Recently, however, for the case of a non-degenerate problem, we proposed a novel, different ansatz: a pseudo-diabatic phase-space surface hopping (PSSH) algorithm\cite{Wu2022}. The basic premise of PSSH is to propagate nuclear motion along phase-space adiabats that naturally incorporate all pseudo-magnetic field effects.  PSSH reduces to FSSH without any spin degrees of freedom and is very simple to rationalize and implement in general. Moreover, in Ref. \citenum{Wu2022}, numerical simulation showed that PSSH  outperforms all FSSH-inspired algorithms for the case of a non-degenerate complex-valued model Hamiltonian when Berry forces are crucial. 

With this background in mind, our goal in this letter is to extend the PSSH algorithm from Ref. \citenum{Wu2022} to a degenerate singlet-triplet ISC model. By comparison with exact data, we will show that the PSSH algorithm algorithm can capture Berry curvature  effects nearly quantitatively  for an interesting model Hamiltonian, outperforming the algorithm published in Ref. \citenum{Bian2022} and with a very simple and generalized algorithm.  As such, in the future, we believe the present algorithm will be the optimal protocol for simulating spin-dependent nonadiabatic dynamics with electronic degeneracy. As a side note, we are also hopeful that the phase-space formulation introduced below can also provide some insights to other MQC methods like MFE and AIMS.

Before concluding this introduction, we emphasize that developing tools to study nonadiabatic transitions with spin degrees of freedom is very important for making progress on one of the most exciting themes today in physical chemistry: the chiral induced spin selectivity (CISS) effect \cite{Naaman2012,Naaman2019}. Recent experiments have demonstrated that, when a current passes through a chiral molecule, that current is often very spin-polarized, and this spin polarization can increase as temperature increases\cite{Fransson2022}.  Thus, understanding nuclear dynamics in the presence of both spin and electronic degrees of freedom would appear crucial for developing a comprehensive model of the CISS effect\cite{Wu2020, Fransson2020, Teh2021, Fay2021, Evers2022}.

\section{\label{sec:method}Method}  
To introduce the PSSH algorithm, let us consider a reasonably generic singlet-triplet ISC Hamiltonian in a spin-diabatic basis of the form $\mathbf{\hat H} =  \mathbf{\hat P}^2 / {2M} + \mathbf{\hat h}_{\rm el}(\mathbf{\hat R})$ with electronic Hamiltonian:
 
\begin{eqnarray} \label{eq:hel}
\mathbf{\hat h}_{\rm el}(\mathbf{\hat R}) = 
\begin{pmatrix}
\epsilon_S & V & Ve^{i\phi} & Ve^{-i\phi}\\
V & \epsilon_T & 0 & 0 \\
Ve^{-i\phi} & 0 & \epsilon_T & 0 \\
Ve^{i\phi} & 0 & 0 & \epsilon_T \\
\end{pmatrix}
\end{eqnarray} 
The singlet spin-diabat $\ket{S}$ with energy $\epsilon_S$ crosses with three triplet spin-diabats $\ket{T_0}, \ket{T_1}, \ket{T_{-1}}$ with energy $\epsilon_T$ and couples with triplets through SOCs with different phases. Here, $(i)$ the diabatic energies $\epsilon_S$ and $\epsilon_T$, $(ii)$ the coupling strength $V$ and $(iii)$ the variable $\phi$ that modulates the complex phase of the SOC are all real-valued and vary with nuclear coordinates $\mathbf{\hat R}$. One would like to simulate nonadiabatic dynamics with such a Hamiltonian while taking into account the fact that $\nabla \phi \ne 0$.

According to PSSH, we apply a basis transformation to cancel the complex-valued phases on the couplings:
\begin{eqnarray}\label{eq:hpseudo}
{\mathbf{\hat{{h'}}}}_{\rm el}(\mathbf{\hat R})  = \mathbf{\Lambda}^\dagger \mathbf{\hat h}_{el}(\mathbf{\hat R})\mathbf{\Lambda} = \begin{pmatrix}
\epsilon_S & V & V  & V \\
V & \epsilon_T & 0 & 0 \\
V & 0 & \epsilon_T & 0 \\
V & 0 & 0 & \epsilon_T \\
\end{pmatrix}  
\end{eqnarray}
where the transformation matrix reads:
\begin{eqnarray}
\mathbf{\Lambda} = \begin{pmatrix} \label{eq:lambda}
1 &0 &0 &0 \\
0 &1 &0 &0 \\
0 &0 &e^{-i\phi} &0 \\
0 &0 &0 & e^{i\phi}
\end{pmatrix}  
\end{eqnarray}
In this new basis, the Hamiltonian is now real-valued and the absolute values of coupling matrix elements are not changed. Note, however, that the basis in Eq. \ref{eq:hpseudo} is no longer diabatic since the electronic states depend on the nuclear coordinates (and so a derivative coupling must emerge). Henceforward, we will refer to this new basis as the  ``pseudo-diabatic'' basis.   The total effective Hamiltonian within the pseudo-diabatic basis then becomes:
\begin{eqnarray}\label{eq:Hpseudo}
\mathbf{\hat H}^{\rm PD}(\mathbf{R}, \mathbf{P}) = \frac {(\mathbf{\hat P} - i\hbar \mathbf{\hat D} (\mathbf{\hat R}) )^2} {2M} + {\mathbf{\hat{{h'}}}}_{\rm el} (\mathbf{\hat R}) 
\end{eqnarray}
where $\mathbf{\hat D} = \mathbf{\Lambda}^\dagger \nabla\mathbf{\Lambda}$ is the derivative coupling operator between pseudo-diabats (we use $\mathbf{\hat D}$ to distinguish this derivative coupling from the derivative coupling $\mathbf{\hat d}$ between adiabats). In  matrix form:
\begin{eqnarray}
\mathbf{\hat D} = \begin{pmatrix}
0 &0 &0 &0 \\
0 &0 &0 &0 \\
0 &0 & -i\nabla \phi &0 \\
0 &0 &0 &i\nabla \phi
\end{pmatrix}  
\end{eqnarray}
At this point, according to PSSH, we apply a Wigner transform to map the quantum mechanical nuclear operators $\mathbf{\hat R}$ and $\mathbf{\hat P}$ to classical variables $\mathbf{R}$ and $\mathbf{P}$. Thereafter, we diagonalize the total phase-space Hamiltonian $\mathbf{\hat H}^{\rm PD}(\mathbf{R}, \mathbf{P})$ in Eq. \ref{eq:Hpseudo}  and we define a new eigenbasis of ``phase-space adiabats'' labeled by $\left\{\ket{\Psi^{\rm PD}_j}\right\}$ 
with eigenenergy $E_j^{\rm PD}$:
\begin{eqnarray}\label{eq:Epd}
\mathbf{\hat H}^{\rm PD}(\mathbf{R}, \mathbf{P}) \ket{\Psi^{\rm PD}_j} = E_j^{\rm PD}(\mathbf{R}, \mathbf{P}) \ket{\Psi^{\rm PD}_j}
\end{eqnarray}
At this point, two points about nomenclature are worth mentioning. First, phase-space adiabats are analogous to the normal position-space adiabats that are obtained from diagonalizing the electronic Hamiltonian $\mathbf{\hat h}_{el}(\mathbf{R})$ in a diabatic basis. For a real-valued Hamiltonian, one can easily see that the pseudo-diabatic derivative coupling matrix $\mathbf{\hat D}$ will be zero and the phase-space adiabats will be identical to the normal position-space adiabats. Second, the phase-space adiabats introduced above are not equivalent to the superadiabats as proposed by Shenvi\cite{Shenvi2009}, the latter of which are obtained from diagonalizing the phase-space Hamitlonian $ \mathbf{\hat H}^{\rm PS}$ in the adiabatic basis  where $ \mathbf{H}^{\rm PS}_{jk} = (\mathbf{P}\delta_{jk} - i\hbar \mathbf{d}_{jk})^2/2M + E_j\delta_{jk}$. The eigenvalues and eigenvectors of both $\mathbf{H}^{\rm PD}$ and $ \mathbf{H}^{\rm PS}$ will both depend on $ \mathbf{R}$ and $\mathbf{P}$, but they will not be the same.

With the background above in mind, we will now outline the PSSH algorithm as appropriate for an ISC event. As inspired by Shenvi's PSSH and following the spirit of Tully's FSSH,  the algorithm starts from sampling a swarm of independent trajectories with a random $\mathbf{R}$ and $\mathbf{P}$ so as to sample the Wigner distribution for a quantum wavepacket.
Each trajectory carries a phase-space quantum amplitude $\mathbf{c}$ and a phase-space adiabat label $n$ that is determined by $\mathbf{c}$. Note that the classical momentum sampled from a quantum wavepacket in a diabatic basis is equivalent to the kinetic momentum $\mathbf{P}^{\rm kinetic} = M\mathbf{\dot R}$. However, PSSH propagates the canonical momentum $\mathbf{P}_n$ (rather than the kinetic momentum).
When working within a pseudo-diabatic basis defined by Eq. \ref{eq:lambda},
the correct relationship between the kinetic and canonical momenta is:
\begin{eqnarray}\label{eq:kinetic}
\mathbf{P}_n = \mathbf{P}^{\rm kinetic}_n + i\hbar \mathbf{D}_{nn}
\end{eqnarray}

At this point, in the spirit of both Hamilton's equations and Tully's algorithm, according to PSSH, the classical variables $\mathbf{R}$ and $\mathbf{P}$ of each trajectory are propagated along a single active phase-space adiabat $n$ according to:
\begin{eqnarray}
     \mathbf{\dot R} = \nabla_{\mathbf{P}} E^{\rm PD}_n 
     = \bra{\Psi_n^{\rm PD}}  \nabla_{\mathbf{P}} \mathbf{\hat H}^{\rm PD} \ket{\Psi_n^{\rm PD}}
\end{eqnarray}
\begin{eqnarray}
     \mathbf{\dot P} = -\nabla_{\mathbf{R}} E^{\rm PD}_n 
     = -\bra{\Psi_n^{\rm PD}}  \nabla_{\mathbf{R}} \mathbf{\hat H}^{\rm PD} \ket{\Psi_n^{\rm PD}}
\end{eqnarray}
The electronic part is treated quantum mechanically and integrated by the Schr\"odinger equation:
\begin{eqnarray}
\dot c_n = -\frac  i {\hbar}  E^{\rm PD}_nc_n - \sum_k    T_{nk} c_k   
\end{eqnarray}
where the time derivative coupling matrix $\mathbf{T}$ can be calculated from the matrix log of the overlap\cite{Jain2016}:
\begin{eqnarray}
T_{nk} = \bra{\Psi_n} \ket{\frac {{\rm d} \Psi_k} {{\rm d }t}} =  \frac {[\log(\mathbf{U})]_{nk}} {\Delta t}
\end{eqnarray}
Here, the overlap matrix $\mathbf{U}$ is defined by $U_{nk} = \bra{\Psi_n(t)}\ket{\Psi_k(t+\Delta t)}$ and $\Delta t$ is the length of discrete time step in the simulation.

At each time step, the trajectory on active phase-space adiabat $n$ has the chance to hop to another phase-space adiabat $k$ with probability:
\begin{eqnarray}
g_{n\to k} = \max\left[2\Delta t \Re\left(T_{nk}\frac{\rho_{kn}}{\rho_{nn}}\right), 0\right]
\end{eqnarray}
\mycomment{where the electronic density matrix elements are $\rho_{kn} = c_kc_n^*$}. After a successful hop from state $n$ to state $k$, the trajectory rescales its momentum along the direction of the position component of the derivative coupling $\mathbf{d}_{nk}^{\mathbf{R},\rm PD}$  between phase-space adiabats $n$ and $k$  to conserve energy \cite{footnote1}, where 
\begin{eqnarray}
\mathbf{d}_{nk}^{\mathbf{R},\rm PD} = \bra{\Psi_n} \nabla_{\mathbf{R}} \ket{\Psi_k} = \frac{\bra{\Psi_n}  \nabla_{\mathbf{R}} \mathbf{\hat H}^{\rm PD} \ket{\Psi_k}} {E^{\rm PD}_k - E^{\rm PD}_n}
\end{eqnarray}
In practice, to satisfy energy conservation, one must search numerically along the direction  $\mathbf{d}_{nk}^{\mathbf{R},\rm PD}$ when rescaling them momentum  because the kinetic energy and potential energy cannot be separated within a phase-space representation; \mycomment{that being said, the energy function is still quadratic in momentum so that one simply picks the closest energy-conserving solution.}  

The description above summarizes the salient features of the PSSH algorithm as compared with FSSH; all other PSSH details are identical with FSSH.  We will now show numerically that this scheme gives a very accurate treatment of ISC dynamics.

Over the last few years, we have diligently attempted to derive a surface hopping ansatz to propagate dynamics for the following model Hamiltonian mimicking a simple singlet-triplet crossing:
\begin{eqnarray}\label{eq:modelH}
\mathbf{\hat h}_{\rm el}  = 
A\begin{pmatrix}
\cos\theta & \frac 1 {\sqrt{3}} \sin \theta & \frac 1 {\sqrt{3}} \sin \theta e^{i\phi} & \frac 1 {\sqrt{3}}\sin \theta e^{-i\phi}\\
\frac 1 {\sqrt{3}}\sin \theta & -\cos\theta & 0 & 0 \\
\frac 1 {\sqrt{3}}\sin \theta e^{-i\phi} & 0 & -\cos\theta & 0 \\
\frac 1 {\sqrt{3}}\sin \theta e^{i\phi} & 0 & 0 & -\cos\theta \\
\end{pmatrix}
\end{eqnarray}

\begin{figure}[H]
    \centering
    \includegraphics[width=\columnwidth]{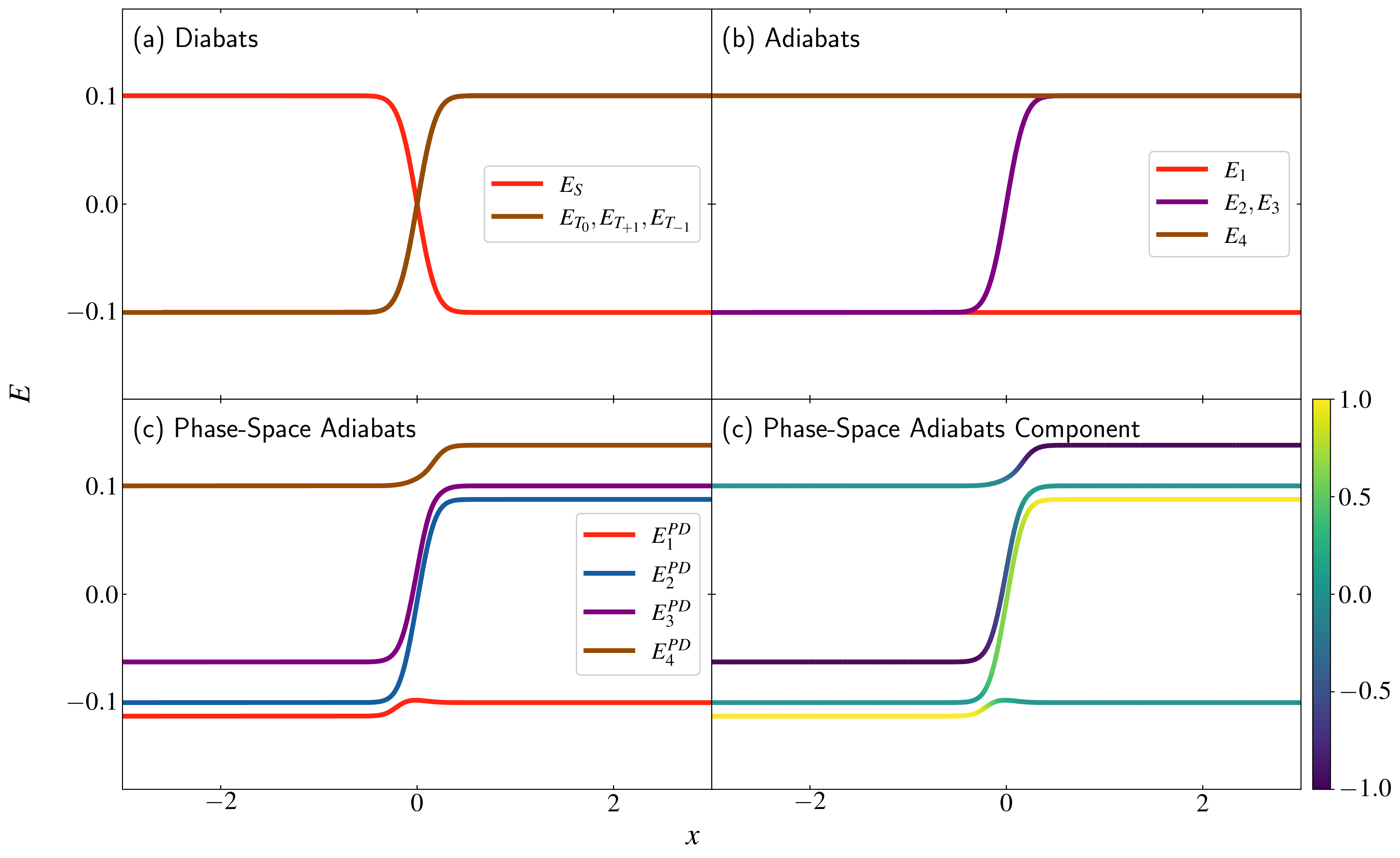}
    \caption{(a) Diabatic, (b) adiabatic and  (c) phase-space adiabatic potentials energy surfaces (PESs) for the singlet-triplet crossing Hamiltonian described in Eq. \ref{eq:modelH} as a function of nuclear coordinate $x$. The phase-space adiabats are calculated for the case $P^{can}_y = 5.0$. Note that, for this Hamiltonian, $P^{can}_y$ is actually a constant of the motion. (d) The color map indicates the spin character $\eta = \bra{\Psi_j^{\rm PD}} \mathbf{\hat \sigma}_z\ket{\Psi_j^{\rm PD}}$ of all the phase-space adiabats. \mycomment{For instance, when $\eta =1$, the phase space adiabat is effectively the $\ket{T_1}$ triplet;  when $\eta =-1$, the phase space adiabat is effectively the $\ket{T_{-1}}$ triplet.  
    This color map allows one to assign spin state heuristically to each phase-space adiabat.} Asymptotically, the phase-space adiabats have a one-to-one mapping to the spin-diabats.}
    \label{fig:PES1}
\end{figure}
 
Here we define $\theta \equiv \frac {\pi} 2 (\erf(Bx) + 1)$ and $\phi \equiv Wy$ and we set $A = 0.10$, $B = 3.0$ and $W = 5.0$ to be constants. 

In Fig. \ref{fig:PES1}, we show a schematic plot of the diabats and  adiabats.
The three triplet diabats are degenerate over all space and cross with the singlet diabat at $x=0$, see Fig. \ref{fig:PES1}(a). The smallest and the largest adiabats $E_1$ and $E_4$ are also flat in the $x$-direction and two degenerate adiabats $E_2$ and $E_3$ connect them in the middle, see Fig. \ref{fig:PES1}(b). The topology of the diabatic and adiabatic PESs are extremely simple for this Hamiltonian.

Now, if one performs a calculation, it is fairly straightforward to observe that, for this problem, the pseudo-diabatic derivative coupling is simply $\hat D_x =0$ and $ \hat D_y = {\rm diag} \begin{pmatrix} 0 & 0 & W & -W \end{pmatrix} $.  Thus, one can calculate phase-space adiabats as functions of nuclear coordinate $x$ (as they are all completely flat in the $y$-direction for the given Hamiltonian in Eq. \ref{eq:modelH}) and $P_y$.  In  \ref{fig:PES1}(c), we plot the phase-space adiabats  with $P_y = W$.  Several interesting features arise upon which we will now elaborate. 

First,  notice that the color map of the spin characters of the phase-space adiabats (as calculated in Fig. \ref{fig:PES1}(d) by calculating $\bra{\Psi_j^{\rm PD}} \mathbf{\hat \sigma}_z\ket{\Psi_j^{\rm PD}}$) shows that this problem is really a complicated crossing of many states with no two-state analogue. 
This is a topologically non-trivial crossing, and the topology of the phase-space adiabats can be very different if one changes $P_y$.

Second, note that the degeneracy of the three adiabats is lifted in 
phase space when $x \to \pm \infty$. The three phase-space adiabats have a one-to-one mapping with three triplet diabats asymptotically given the fact that $V \to 0$ as $x \to \pm \infty$ and the total Hamiltonian matrix in Eq. \ref{eq:Hpseudo} is diagonal. For instance, when $x \to \infty$, $\ket{\Psi_3^{\rm PD}}$ should map to $\ket{T_0}$, and $\ket{\Psi_2^{\rm PD}}, \ket{\Psi_4^{\rm PD}}$ should map to $\ket{T_{+1}},\ket{T_{-1}}$ correspondingly. This mapping and breaking of degeneracy is also made clear by Fig. \ref{fig:PES1}(d).

Third, consider an incoming trajectory on the singlet diabat $\ket{S}$ from the left with $\mathbf{P}= (P_x, P_y)$ and assume $P_y > 0$. In  phase space, the asymptotic energy barrier to transmit to the right on the upper surfaces should be increased by $(2P_y W + W^2)/2M$ on $\ket{\Psi_4^{\rm PD}}$ (or $\ket{T_{-1}}$) and decreased by $(-2P_y W + W^2)/2M $ on $\ket{\Psi_2^{\rm PD}}$ (or $\ket{T_{+1}}$). Thus,  for the transmitting trajectories, one of the real physical observables -- the kinetic momentum $\mathbf{P}^{\rm kinetic}$ -- will be very different according to Eq. \ref{eq:kinetic} for trajectories emerging on different phase-space adiabats , i.e. the nuclear motion will bend in different directions depending on the corresponding electronic state. Intuitively, without running any PSSH simulations, these three facts can  explain a lot of the exotic bending and bifurcation phenomena summarized in  Table. 1 in Ref. \citenum{Bian2022}.

To quantitatively assess the performance of the present PSSH algorithm, we have simulated scattering dynamics for the Hamiltonian in Eq.\ref{eq:modelH} and compared the results for transmission and reflection probabilities according to both exact wavepacket dynamics and Tully's standard FSSH algorithm.  
For the wavepacket dynamics, the system is initialized as a Gaussian wavepacket with the form:
\begin{eqnarray}\label{eq:psi0}
    \ket{\Psi_0 (\mathbf{R})} =  \exp \left(-\frac {({\mathbf{R}} - {\mathbf{R_0}})^2} {\mathbf{\sigma}^2} +   \frac i \hbar {{\mathbf{P}}_0 \cdot {\mathbf{R}}}   \right) \ket{\phi_i}  
\end{eqnarray}
We set the initial position $\mathbf{R_0} = (-4,0)$, the wavepacket width is set to $\sigma_x = \sigma_y = 1$, the nuclear mass $M = 1000$. We chose an initial momentum $\mathbf{P_0} = (P^{\rm init}_x,P^{\rm init}_y)$ with $P^{\rm init}_x = P^{\rm init}_y$ and we scan over the range $P^{\rm init}_x = 8$ to $24$; all parameters are in atomic units. The initial wavepacket is set to be on one of the pure spin-diabats. Exact wavepacket dynamics are propagated on a two dimensional grid by the split-operator method with the fast Fourier transform algorithm\cite{Kosloff1983}.

For the PSSH data, all results were generated following the method as described above. The initial active phase-space adiabat of each trajectory was randomly assigned according to the quantum amplitude in the phase-space adiabatic basis. For PSSH, no velocity reversal (upon a frustrated hop) was implemented; for standard FSSH results, velocity reversal was administered in the spirit of Truhlar's ``$\nabla V$'' approach\cite{Jasper2003, footnote2} .

For both surface hopping methods, $2 \times 10^3$ trajectories were sampled according to the Wigner distribution of Eq. \ref{eq:psi0}.

\begin{figure}[H]
    \centering
    \includegraphics[width=1\columnwidth]{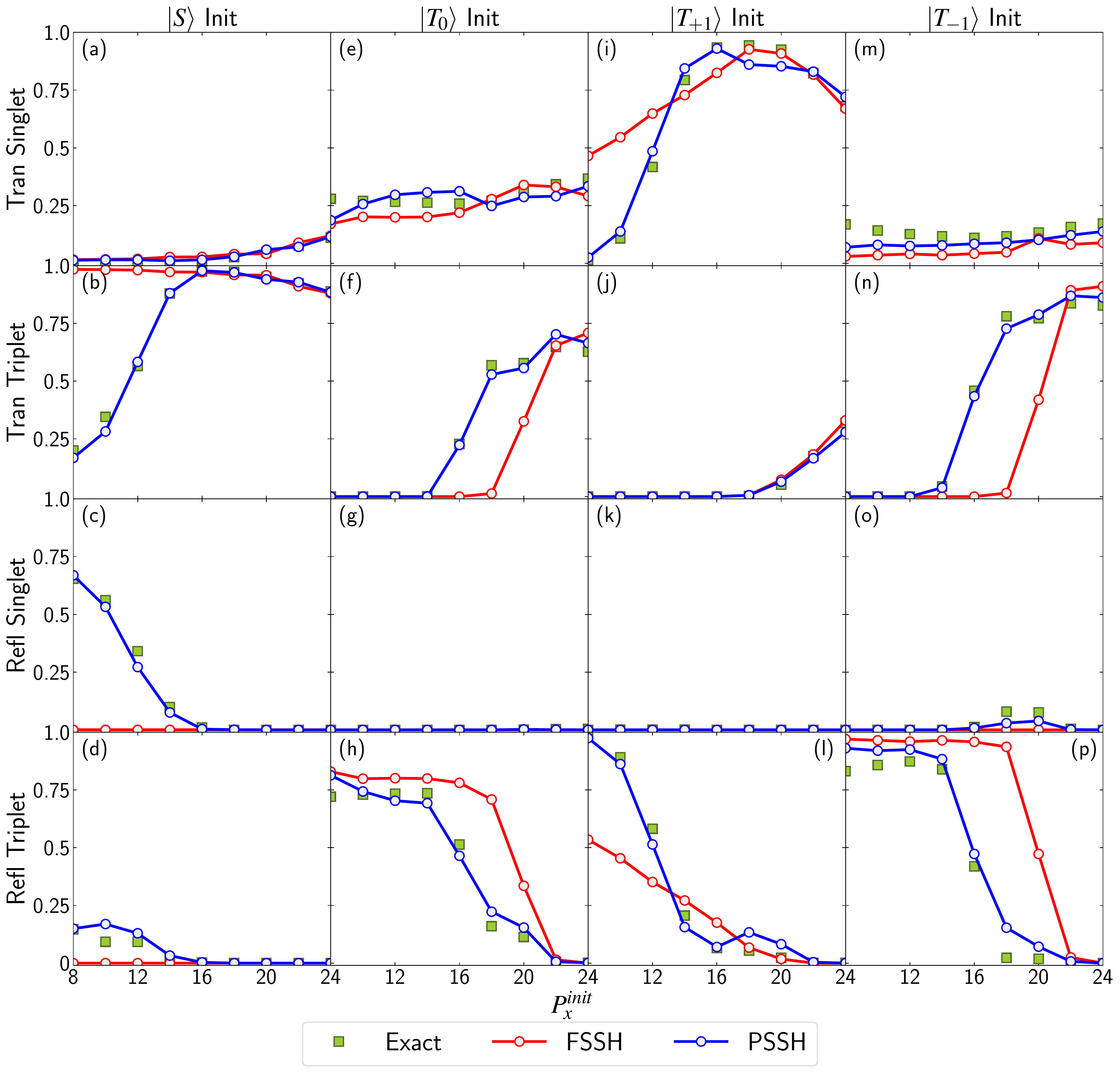}
    \caption{The results for transmission and reflection populations according to exact dynamics, standard surface hopping and phase-space surface hopping algorithms. The systems are initialized on  (a) - (d) the singlet spin-diabat $\ket{S}$,  (e) - (h) the triplet $\ket{T_0}$, (i) - (l) the triplet $\ket{T_{+1}}$ and (m) - (p) the triplet $\ket{T_{-1}}$.
    The PSSH method shows strong agreement with the exact results, where  standard FSSH often is very incorrect. 
    }
    \label{fig:A010ang1}
\end{figure}

Simulation results are presented in Fig. \ref{fig:A010ang1}. 
We find that Berry curvature effects lead to a change in the effective energy barrier for transmission on the different surfaces. 
For instance, according to exact wavepacket dynamics, a large population is reflected for the case that the system begins on the upper singlet at low momentum ($P_x^{\rm init} < 16$), see Fig. \ref{fig:A010ang1}(a) - (d). Such reflection is completely missing from the standard FSSH method. 
Similarly, for the cases that dynamics begin on a lower triplet, the energy barrier for transmission is altered by the Berry curvature. The standard FSSH routinely predicts qualitatively wrong results, while the pseudo-diabatic PSSH method matches with exact results almost quantitatively in all cases (as well and usually better than the algorithm presented in Ref. \citenum{Bian2022}). \mycomment{Clearly, the PSSH algorithm outperforms the standard FSSH algorithm in all cases because the FSSH algorithm completely neglects the contribution of Berry curvature effects, whereas the PSSH includes such effects explicitly.}
More results for systems beginning with different initial conditions can be found in the supporting information.

The strong results above have been achieved following a simple surface hopping protocol (as opposed to the approach in Ref. \citenum{Bian2022}) and make clear that, if possible, a pseudo-diabatic PSSH formulation is the the best protocol for incorporating Berry curvature effects into spin-dependent MQC dynamics. That being said, the algorithm described above is still not fully tested yet; beyond connecting the algorithm to more rigorous theories of nonadidabatic dynamics\cite{littlejohn1991Geometric,weigert1993Diagonalization}, several practical points must be explored before we can begin running first principles simulations. 

The first and most important question that must be addressed is the question of how to choose a basis for the PSSH pseudo-diabats. After all, for any Hamiltonian, there is no one fixed (unique) diabatic basis. \mycomment{For instance, we have constructed the Hamiltonian in Eq. \ref{eq:hel} in an arbitrary lab frame with fixed $x$,$y$, and $z$ directions;} if one were to apply a fixed global change of basis matrix and then boost the basis functions for the final Hamiltonian, one would arrive at a different set of pseudo-diabats (and therefore different PSSH dynamics). Vice versa, given that one can change basis to render the entire Hamiltonian in Eq. \ref{eq:hel} real-valued\cite{Mead1979d}, one might also conclude that there is no reason to choose pseudo-diabats that differ from true diabats, and therefore  PSSH dynamics would be exactly equivalent to FSSH dynamics. In short, the present PSSH algorithm sorely depends the choice of diabats, and in the future, it will be essential to pick a good diabatic starting frame. The essential features of this frame are not yet known, but judging from the model presented here, ideally one would like a frame for which the norms of the derivative of SOC matrix elements  ($\abs{\nabla H_{\rm SOC}}$) are small.

Second, so far we have considered an isolated singlet-triplet crossing. In the presence of an external magnetic field $\mathbf{B}$,  the electronic Hamiltonian will break some of the  degeneracy through the Zeeman interaction:
\begin{eqnarray} \label{eq:helB}
\mathbf{\hat h}_{\rm el}(\mathbf{\hat R}) = 
\begin{pmatrix}
\epsilon_S & V & Ve^{i\phi} & Ve^{-i\phi}\\
V & \epsilon_T   & B_x -iB_y & B_x +iB_y \\
Ve^{-i\phi} & B_x + iB_y  & \epsilon_T + B_z & 0 \\
Ve^{i\phi} & B_x - iB_y & 0 & \epsilon_T - B_z\\
\end{pmatrix}
\end{eqnarray} 
In such a case, one must ask: what is the correct basis for boosting the electronic frame and running PSSH? The answer may not appear obvious, but one path forward would be to orient the basis functions so that $\hat z$ is collinear with $\mathbf{B}$ (so that ${B_x} = {B_y} = 0$). In such a case, the problem is rendered completely  equivalent to the problem above. Thus, if one can, in general, define an optimal basis for the singlet-triplet problem without a magnetic field present, one can also likely construct an optimal basis for the singlet-triplet problem with a magnetic field present.

Third, simulating spin-dependent nonadiabatic nuclear-electronic dynamics in the condensed phases raises a host of questions that must be addressed. In a previous study using non-equilibrium Fermi's golden rule\cite{Chandran2022}, we found that spin polarization survived for some transient period of time for a simple spin-boson model with a complex-valued coupling. In the future, it will be important to test whether the PSSH algorithm can recover such transient effects in complicated environments with friction; it will also be important to explore if/how PSSH recovers detailed balance in the condensed phase.  Lastly, the question of decoherence\cite{Bittner1995,Subotnik2011} within the PSSH algorithm must be related to the ever fascinating question of spin-lattice relaxation, which again gives us a strong motivation to further explore spin-dependent nonadiabatic dynamics with a surface hopping formalism.

In conclusion, we have proposed a simple and general pseudo-diabatic phase-space surface hopping method for simulating spin-dependent nonadiabatic dynamics with electronic degeneracy as relevant to singlet-triplet ISC dynamics. For a simple model system, the algorithm shows significant improvements over the standard position-space surface hopping and captures  Berry curvature effects while keeping the spirit of Tully's original FSSH algorithm (and Shenvi's insight into superadiabats \cite{Shenvi2009}).
Looking forward, \mycomment{there are a host of intriguing chemical reactions that can be altered by magnetic fields\cite{Steiner1989}. Given the fact that the nuclear Berry curvature usually acts like a pseudo-magnetic field, it is likely nuclear Berry curvature effects may be paramount in such magnetic chemical systems as well.}
Furthermore, there are a myriad of photochemical problems occurring on triplet states where ISC plays a key role\cite{Molander2020,Nicewicz2016}. We are hopeful this method (and variants thereof) will soon be applied to model such systems with {\em ab initio} electronic structure. 

\begin{suppinfo}
Additional numerical results can be found in the supporting information, including   several different Hamiltonian parameters with different initial conditions. We also provide a comparison between the current PSSH method versus FSSH and quasi-diabatic Berry force method in Ref.\citenum{Bian2022}.
\end{suppinfo}

\section{Acknowledgements}
This material is based on the work supported by the National
Science Foundation under Grant No. CHE-2102402.

\bibliography{main.bib}

\begin{mcitethebibliography}{59}
\bibitem[foo()]{footnote2}
We also computed results for Tully's FSSH algorithm without any velocity
  reversal, and the results were equally disappointing.\relax
\mciteBstWouldAddEndPunctfalse
\mciteSetBstMidEndSepPunct{\mcitedefaultmidpunct}
{}{\mcitedefaultseppunct}\relax
\EndOfBibitem
\end{mcitethebibliography}
\end{document}